\documentclass[12pt,a4paper]{article}

\usepackage[tbtags]{amsmath}

\advance\textheight by \topskip
\let\oldappendix=\appendix
\let\oldsection=\section
\renewcommand{\appendix}{\oldappendix%
\def\theequation{\Alph{section}.\arabic{equation}}%
\renewcommand{\section}{\setcounter{equation}{0}\oldsection}}

\newcommand{\bibeprint}[1]{\texttt{#1}}
\newcommand{\bibbtitle}[1]{\textit{``#1''}}
\newcommand{\bibtitle}[1]{ \bibbtitle{#1},}
\newcommand{\bibextra}[1]{ [#1]}
\newcommand{\bibstart}{}

\newcommand{\I}{i}

\newcommand{\sfrac}[2]{{\textstyle\frac{#1}{#2}}}
\newcommand{\mcirc}[2][2]{{\overset{\circ}{m}\vphantom{m}_{#2}^{#1}}}
\newcommand{\mmean}[1][2]{\mcirc[#1]{8}}
\newcommand{\msplit}[1][2]{\mcirc[#1]{\Delta}}
\newcommand{\mtop}[1][2]{\mcirc[#1]{0}}
\newcommand{\tad}[1]{\Delta_{#1}}
\newcommand{\tsum}{\mathop{\textstyle\sum}}
\newcommand{\cder}{D}
\newcommand{\decay}{f}
\newcommand{\coeffv}[2]{v_{#1}^{(#2)}}
\newcommand{\cbeta}[2]{\beta_{#1}^{(#2)}}
\newcommand{\cvtwid}[2]{\tilde{v}_{#1}^{(#2)}}

\newcommand{\MeV}{\,\mathrm{MeV}}
\newcommand{\GeV}{\,\mathrm{GeV}}
\newcommand{\Lagr}{\mathcal{L}}

\begin{document}

\hfill 

\hfill 

\bigskip\bigskip

\begin{center}

{{\Large\bf  $\mbox{\boldmath$\eta$}$-$\mbox{\boldmath$\eta'$}$ mixing in\\
$\mbox{\boldmath$U(3)$}$ chiral perturbation theory 
\footnote{Work supported in part by the DFG}}}

\end{center}

\vspace{.4in}

\begin{center}
{\large N. Beisert\footnote{email: nbeisert@physik.tu-muenchen.de}
 and  B. Borasoy\footnote{email: borasoy@physik.tu-muenchen.de}}

\bigskip

\bigskip

Physik Department\\
Technische Universit{\"a}t M{\"u}nchen\\
D-85747 Garching, Germany \\

\vspace{.2in}

\end{center}

\vspace{.7in}

\thispagestyle{empty} 

\begin{abstract}
We investigate $\eta$-$\eta'$ mixing in infrared regularized
$U(3)$ chiral perturbation theory by calculating the $\eta$ and $\eta'$ 
masses up to one-loop order. From this analysis it becomes obvious
that even at leading order $\eta$-$\eta'$ mixing does not obey the usually 
assumed one-mixing angle scheme if large $N_c$ counting rules are not employed.
\end{abstract}

\vfill

\section{Introduction}\label{sec:intro}

The $\eta$-$\eta'$ mixing has been the subject of many investigations,
see e.g. \cite{AF}-\cite{H-S}. Both particles can be described as mixtures of
the octet component $\eta_8$ and its singlet counterpart $\eta_0$.
The $\eta_8$ which is a member of the octet of the pseudoscalar mesons
($\pi, K, \eta_8$) differs from the singlet $\eta_0$ in a substantial way: it
is a Goldstone boson whose mass vanishes in the limit of zero quark masses
while the $\eta_0$ is not due to the axial $U(1)$ anomaly.

Phenomenologically, however, the situation for the $\eta$-$\eta'$ mixing still
remains to be settled. Most of the investigations on this subject introduce one
single mixing angle and extract a value from different kinds of data. 
These are, e.g., the anomalous $\eta ,\eta'$ decays, $\eta ,\eta'
\rightarrow \gamma \gamma $ \cite{AF,VH}, decays of $J/\Psi$ 
\cite{GK,BES1,BFT}, electromagnetic decays of vector and pseudoscalar mesons
\cite{BES2}, only to name a few. The values obtained in these investigations
range from $-13^\circ$ \cite{BES2} to $-22^\circ$ \cite{VH}.
On the other hand, the Gell-Mann--Okubo mass formula for the pseudoscalar
mesons yields a mixing angle of $-10^\circ$ \cite{DGH}.

More recently, a two-mixing angle scheme has been proposed by Kaiser
and Leutwyler 
\cite{L,KL1,KL2} for the calculation of the pseudoscalar decay constants in
large $N_c$ chiral perturbation theory. The two angle scenario has been adopted
in a phenomenological analysis on the two-photon decay widths of the $\eta$ and
$\eta'$, the $\eta \gamma$ and $\eta' \gamma$ transition form factors,
radiative $J/\Psi$ decays, as well as on the  decay constants of the
pseudoscalar mesons \cite{FK1,FK2}.
The authors observe that within their phenomenological approach the assumption
of one mixing angle is not in agreement with experiment whereas the two-mixing
angle scheme leads to a very good description of the data.
These two different mixing angles have been interpreted as one 
energy-dependent 
$\eta$-$\eta'$ mixing angle in \cite{EF} where electromagnetic couplings
between lowest-lying vector and pseudoscalar mesons were studied.
As pointed out in these investigations the analysis with two different mixing
angles leads to a more coherent picture than the canonical treatment with a
single angle.
In particular, the calculation of the pseudoscalar decay constants within the
framework of large $N_c$ chiral perturbation theory
${\it requires}$ two different mixing angles \cite{L}.
(A similar investigation was performed in \cite{H-S} but with a different
parametrization.)

Recently, it has been shown in \cite{BW} that the $\eta'$ can be
included in a systematic way in chiral perturbation theory without employing
1/$N_c$ counting rules. The loop integrals are evaluated using infrared
regularization, which preserves Lorentz and chiral symmetry \cite{BL}.
However, in \cite{BW} 
it was assumed that the $\eta$-$\eta'$ mixing follows 
at lowest order in symmetry breaking the one-mixing angle pattern,
i.e. the mixing is described only by one mixing
angle and its value was assumed to be $-20^\circ$.

The purpose of this work is to critically investigate $\eta$-$\eta'$ mixing up 
to one-loop
order in infrared regularized $U(3)$ chiral perturbation theory which provides
a systematic counting scheme. Within this approach loops start contributing at
next-to-leading order while they are a next-to-next-to-leading order effect in
large $N_c$ chiral perturbation theory.

We start in the next section by presenting the effective Lagrangian and 
$\eta$-$\eta'$ mixing at lowest order. 
The next-to-leading order calculation
within this counting scheme including one-loop diagrams
is presented in Section~\ref{sec:loopmass}. 
We also compare this approach with a scheme that takes only loops with
Goldstone bosons into account omitting any propagation of an
$\eta'$ inside the loop.
Section~\ref{sec:results} contains our results and 
we conclude with a summary in Section~\ref{sec:conclusions}.

\section{$\mbox{\boldmath$\eta$}$-$\mbox{\boldmath$\eta'$}$ mixing at
leading order}
\label{sec:leadmix}

In this section, we present $\eta$-$\eta'$ mixing at lowest order in the
framework of infrared regularized $U(3)$ chiral perturbation theory.
Note that we do not make use of 1/$N_c$ counting rules.
The effective Lagrangian for the pseudoscalar meson nonet
($\pi, K, \eta_8, \eta_0$) reads up to second order in the derivative expansion
\cite{KL1,KL2,BW}
\footnote{If one prefers, one can transform the $V_5$ term away. Here
we keep it for completeness.}
\begin{eqnarray}  \label{eq:mes1}
\Lagr^{(0+2)} &=& - V_0 
+V_1 \langle \cder_{\mu} U^{\dagger} \cder^{\mu}U \rangle  
+V_2 \langle U^\dagger \chi+\chi^\dagger U\rangle
+i V_3 \langle U^\dagger \chi-\chi^\dagger U\rangle
\nonumber\\&&\mathord{}
+V_4 \langle U^\dagger\cder^{\mu} U \rangle \langle U^\dagger\cder_{\mu} U \rangle
+i V_5 \cder^\mu\theta \langle U^\dagger\cder_{\mu} U \rangle
+V_6 \cder^\mu\theta\cder_\mu\theta,
\end{eqnarray}
where $U$ is a unitary $3 \times 3$ matrix containing the Goldstone boson
octet ($\pi, K, \eta_8$) and the $\eta'$. 
Its dependence on $\eta_8$ and $\eta_0$ is given by
\begin{equation}
U=\exp\bigl(\mathrm{diag}(1,1,-2)\cdot i\eta_8/ \sqrt{3}\decay
+i\sqrt{2}\eta_0/\sqrt{3}\decay+\ldots\bigr).
\end{equation}
The expression $\langle \ldots \rangle$ denotes 
the trace in flavor space, $\decay$ is the pion decay constant in the chiral limit
and the quark mass matrix $\mathcal{M} = \mbox{diag}(m_u,m_d,m_s)$
enters in the combination  $\chi  = 2 B \mathcal{M} $
with $B = - \langle  0 | \bar{q} q | 0\rangle/ \decay^2$ being the order
parameter of the spontaneous symmetry violation.
The external field $\theta$ is the QCD vacuum angle, which will be
set to zero throughout this discussion.
The covariant derivatives are defined by
\begin{eqnarray}
\cder_{\mu} U  &=&  \partial_{\mu} U - i ( v_{\mu} + \tilde{a}_{\mu}) U
                     + i U ( v_{\mu} - \tilde{a}_{\mu})   \nonumber \\
\cder_{\mu} \theta  & = &  \frac{\sqrt{6\lambda}}{\decay}\partial_{\mu} \theta 
        + 2 \langle \tilde a_{\mu} \rangle.
\end{eqnarray}
They are defined in such a way, that 
all the dependence on the running scale of QCD
due to the anomalous dimension of the singlet axial current
$A_\mu^0 = \frac{1}{2} \bar{q} \gamma_\mu \gamma_5 q$
is absorbed into the prefactor $\sqrt{\lambda}$, cf. \cite{BW} for details.
Due to its scale dependence, $\sqrt{\lambda}$ cannot be determined from experiment, and
all quantities involving it are unphysical.
The axial-vector connection $\tilde a_\mu$ is defined as 
\begin{equation}\label{eq:scalea0}
\tilde a_\mu=a_\mu + \frac{\sqrt{6\lambda}-f}{3f} \langle a_\mu \rangle.
\end{equation}
which is the scale independent combination of the octet and singlet parts of the
external axial-vector field $a_\mu$.

For $\theta=0$ the coefficients $V_i$ are functions of 
$\eta_0$,
$V_i(\eta_0/\decay)$,
and can be expanded in terms of this variable. At a given order of
derivatives of the meson fields $U$ and insertions of the quark mass matrix 
$\mathcal{M}$ one obtains an infinite string of increasing powers of 
$\eta_0$ with couplings which are not fixed by chiral symmetry.
Parity conservation implies that the $V_i$ are all even functions
of $\eta_0$ except $V_3$, which is odd, and
$V_1(0) = V_2(0) = V_1(0)-3V_4(0)=\frac{1}{4}\decay^2$
 gives the correct  normalizaton
for the quadratic terms of the mesons.
The potentials $V_i$ are expanded in the singlet field $\eta_0$ 
\begin{eqnarray}\label{eq:vexpand}
V_i\Big[\frac{\eta_0}{\decay}\Big] &=& \coeffv{i}{0} + \coeffv{i}{2} 
\frac{\eta_0^2}{\decay^2} +
\coeffv{i}{4} \frac{\eta_0^4}{\decay^4} + \ldots
\qquad \mbox{for} \quad i= 0,1,2,4,5,6 \nonumber \\
V_3\Big[\frac{\eta_0}{\decay}\Big] &=& \coeffv{3}{1} \frac{\eta_0}{\decay} + 
\coeffv{3}{3} \frac{\eta_0^3}{\decay^3}
+ \ldots \quad 
\end{eqnarray}
with expansion coefficients $\coeffv{i}{j}$ to be determined phenomenologically.
In the present investigation we work in the isospin limit
$m_u=m_d=\hat{m}$ and, therefore, only $\eta$-$\eta'$ mixing occurs.
One observes terms quadratic in the meson fields that contain the factor
$\eta_0 \eta_8$. Such terms arise from the explicitly chiral symmetry breaking
operators $V_2 \langle U^\dagger \chi+\chi^\dagger U\rangle
+i V_3 \langle U^\dagger \chi-\chi^\dagger U\rangle$
and read
\begin{equation} \label{eq:mix1}
- \frac{8\sqrt{2}}{3 \decay^2} 
\big( \sfrac{1}{4}\decay^2-\sfrac{1}{2}\sqrt{6}\coeffv{3}{1} \big) B (\hat{m}
  - m_s) \eta_0 \eta_8 .  
\end{equation}
However, these are not the only $\eta_0$-$\eta_8$ mixing terms arising at
second chiral order. Terms from $\Lagr^{(4)}$, 
the Lagrangian at fourth chiral order,
which is presented in App. \ref{sec:fourth}, will also contribute to the mass matrix
at second chiral order. This can be seen as follows.
Consider the terms in $\Lagr^{(4)}$ with one or two derivatives of the
singlet field and an
insertion of the quark mass matrix $\chi$.
They are given by \cite{BW,H-S2}
\begin{eqnarray}  \label{eq:higher}
\Lagr^{(4)} &=& \ldots  
+  \frac{2}{3 \decay^2} (3 \beta_{4} + \beta_{5} - 9 \beta_{17} + 3 \beta_{18} )
e^{-i \sqrt{6} \eta_0/(3\decay)}  
\cder_{\mu} \eta_0 \cder^{\mu} \eta_0 
 \langle \hat U^{\dagger} \chi \rangle  + h.c.
\nonumber\\&&\mathord{}
+ \frac{i\sqrt{6}}{3\decay} 
(2 \beta_{5} + 3 \beta_{18})
e^{-i \sqrt{6} \eta_0/(3\decay)} \cder_{\mu} \eta_0 
\langle \cder^{\mu}\hat U^\dagger \chi \rangle +h.c.
+\ldots
\end{eqnarray}
where we have kept the notation from \cite{BW} and $\hat U=(\det U)^{-1/3} U$
contains only Goldstone boson fields.
The $\beta_i$ are functions of the singlet field $\eta_0$ and can be expanded
as in Eq. \eqref{eq:vexpand}.
They contribute to the part
of the effective Lagrangian quadratic in the $\eta_8$ and $\eta_0$
which has the generic form
\begin{equation} \label{eq:kinetic}
\Lagr = \sfrac{1}{2} \partial_\mu \eta_i \big(\delta_{ij} + K_{ij}^{(2)} \big)
 \partial^\mu \eta_j - \sfrac{1}{2} \eta_i \big( M_{ij}^{(0)} + M_{ij}^{(2)}\big)
  \eta_j  , \qquad i,j=0,8 
\end{equation}
where the superscripts for the matrices $K$ and $M$ denote the chiral power.
(We restrict ourselves to the $\eta$-$\eta'$ system since pions and kaons
decouple in the isospin limit.)
Choosing $K$ and $M$ in a symmetric form one obtains from
Eqs. \eqref{eq:mes1} and \eqref{eq:higher} the non-vanishing coefficients
\begin{eqnarray}\label{eq:masskintwo}
M_{00}^{(0)} &=& \mtop, \nonumber\\
M_{88}^{(2)} &=& \mmean+\sfrac{1}{2}\msplit,\nonumber\\
M_{08}^{(2)} &=& -2\sqrt{2}\cvtwid{2}{1}\msplit\big/\decay^2,\nonumber\\
M_{00}^{(2)} &=& 4\cvtwid{2}{2}\mmean\big/\decay^2,\nonumber\\
K_{08}^{(2)} &=& -4 \sqrt{2}\beta_{5,18}\msplit\big/\decay^2 ,\nonumber\\
K_{00}^{(2)} &=& 8\beta_{4,5,17,18}\mmean\big/\decay^2.
\end{eqnarray}
Here we have made the following abbreviations for combinations 
of constants that repeatedly occur
\begin{eqnarray}\label{eq:abbrev}
\mtop&=&\frac{2\coeffv{0}{2}}{\decay^2},
\nonumber\\
\mmean&=&\sfrac{2}{3}B(2\hat m+m_s),
\nonumber\\
\msplit&=&\sfrac{4}{3}B(m_s-\hat m),
\nonumber\\
\cvtwid{2}{1}&=&\sfrac{1}{4}\decay^2-\sfrac{1}{2}\sqrt{6}\coeffv{3}{1},
\nonumber\\
\cvtwid{2}{2}&=&\sfrac{1}{4}\decay^2-\sqrt{6}\coeffv{3}{1}-3\coeffv{2}{2},
\nonumber\\
\beta_{5,18}&=&\cbeta{5}{0}+\sfrac{3}{2}\cbeta{18}{0},
\nonumber\\
\beta_{4,5,17,18}&=&3\cbeta{4}{0}+\cbeta{5}{0}-9\cbeta{17}{0}+3\cbeta{18}{0}.
\end{eqnarray}
The mass of the $\eta'$ in the chiral limit is denoted by $\mtop[]$, 
$\mmean$ is the mean mass squared of the octet, 
$\msplit$ describes the mass splitting of the octet.
Both combinations $\cvtwid{2}{1}$ and $\cvtwid{2}{2}$ 
approach $\frac{1}{4}\decay^2$ 
in the large $N_c$ limit.
The scale dependence of the renormalized $\beta_i$ parameters cancels 
in the combinations $\beta_{5,18}$ and $\beta_{4,5,17,18}$.

The wave functions must be renormalized in order to acquire the
canonical form for the Lagrangian
\begin{equation} 
\Lagr = 
\sfrac{1}{2} \partial_\mu \eta\partial^\mu \eta
+\sfrac{1}{2} \partial_\mu \eta'\partial^\mu \eta'
-\sfrac{1}{2} \mcirc{\eta}\eta^2
-\sfrac{1}{2} \mcirc{\eta'}\eta^{\prime 2}.
\end{equation}
To second order this is achieved by the transformation
$(\eta_8,\eta_0)^T=(1+R_0^{(2)})(\eta,\eta')^T$
with
\begin{equation}\label{eq:waverenormlag}
1+R_0^{(2)}=\left(\begin{array}{cc}
1&M_{08}^{(2)}\big/M_{00}^{(0)}-K_{08}^{(2)}\\[3pt]
-M_{08}^{(2)}\big/M_{00}^{(0)}
&1-\frac{1}{2} K_{00}^{(2)}\end{array}\right).
\end{equation}
The off-diagonal elements of this transformation
describe mixing between the fields $\eta$ and $\eta'$. 
Even in leading order the two off-diagonal 
elements are different  in contradistinction to
large $N_c$ chiral perturbation theory 
(cf. \cite{KL1,H-S2}),
where the sum of both off-diagonal elements vanishes
in leading order. There the term $K_{08}^{(2)}$ is 
of higher order than $M_{08}^{(2)}/M_{00}^{(0)}$
and it is justified to use just one mixing angle.
In our approach we have two different off-diagonal elements
which leads directly to two different mixing angles.

The two off-diagonal elements are not the two mixing angles
occurring in the pseudoscalar decay constants
of the $\eta$ and the $\eta'$, although they are closely 
related. In leading order the 
$\eta_i$ fields couple to $\tilde a^\mu_i$ with
strength $\decay$ for $i=8$ and 
$\decay_0= \sqrt{6 \lambda} (1 + 6 \coeffv{5}{0}/ \decay^2)$ for $i=0$, 
see Eq.~\eqref{eq:decay0}.
After the transformation to $\eta,\eta'$ fields
the coupling matrix can be written as $\mbox{diag}(\decay,\decay_0)(1+R_0^{(2)})$.
We will see later that there are also loop
corrections to the coupling matrix in second chiral order,
however, $R_0^{(2)}$ involves two amplitudes and two mixing angles
already at tree level.
For the full results, see Eqs. \eqref{eq:axcouptwo} and \eqref{eq:axangletwo}.

Note also, that $(1+R_0^{(2)})$ in Eq. \eqref{eq:waverenormlag} is not the complete 
wave function renormalization to second chiral order; there
are corrections from loops and LECs which are presented in the next
section. However, they do not affect the masses at this order and could be
dropped so far. The full matrix is given in Eq. \eqref{eq:lastmix}

After the transformation the masses can be read off from the
Lagrangian, they are given by
$\mcirc{\eta}=M_{88}^{(2)}$ and
$\mcirc{\eta'}=M_{00}^{(0)}+M_{00}^{(2)}-M_{00}^{(0)}K_{00}^{(2)}$.
Expressed in $U(3)$ parameters the masses at second chiral order are
\begin{eqnarray}\label{eq:masstwo}
\mcirc{\pi} &=& \mmean-\sfrac{1}{2}\msplit \nonumber \\
\mcirc{K} &=& \mmean+\sfrac{1}{4}\msplit \nonumber \\
\mcirc{\eta} &=& \mmean+\sfrac{1}{2}\msplit\nonumber \\
\mcirc{\eta'} &=& 
\mtop + \frac{4}{\decay^2}\mmean
\big(\cvtwid{2}{2}- 2\mtop\beta_{4,5,17,18} \big)
\end{eqnarray}
where we have included the $\pi$ and $K$ masses for completeness.
Note that the Gell-Mann--Okubo mass relation is satisfied in leading
order.

\section{Inclusion of loops}\label{sec:loopmass}

We proceed by investigating $\eta$-$\eta'$ mixing 
in the calculation of the $\eta$ and $\eta'$ masses at next-to-leading 
order. To this order contributions both from one-loop graphs and higher
order contact terms must be taken into account.
The fourth order Lagrangian is given by 
\begin{equation}\label{eq:l4}
\Lagr^{(4)}=\tsum\nolimits_k \beta_k O_k,
\end{equation}
where the fourth order operators are given in App. \ref{sec:fourth}.
In the present work
the contributing operators $O_k$ are those with 
$k=0,\ldots,8,12,13,14,15,25,26$.
We have decided to include the $\beta_0$ term, although
there is a Cayley-Hamilton matrix identity that enables one to 
remove the term leading to modified coefficients 
$\beta_i$, $i=1,2,3,13,14,15,16$ \cite{H-S2}.
(It is actually more convenient
to eliminate one of the OZI violating terms $\beta_{14}$, 
$\beta_{15}$ or $\beta_{16}$, see \cite{H-S4}.)
Here we do not make use of the Cayley-Hamilton identity
and keep all couplings in order to present the most general expressions
in terms of these parameters. One can then drop one of the 
$\beta_i$ involved in the Cayley-Hamilton identity
at any stage of the calculation.
Furthermore, one-loop diagrams from the Lagrangian in Eq. \eqref{eq:mes1}
contribute at this order. It is crucial to employ infrared regularization in
the evaluation of the loop diagrams if one does not implement large $N_c$
counting rules. Otherwise, the inclusion of $\eta'$ loops would spoil the
counting scheme and in general higher loops with an arbitrary number of
$\eta'$-propagators will contribute to lower chiral orders. This is similar to
the situation in the relativistic framework of dimensionally regularized baryon
chiral perturbation theory. Using infrared regularization allows for a chiral
counting scheme while preserving chiral invariance \cite{BL}.
The loop diagrams are usually divergent and must be renormalized
by counterterms of arbitrarily high order.
This cannot be done in practice and one neglects these counterterm
polynomials \cite{BL}.
We will proceed in a similar way,  restricting ourselves to the calculation of the chiral logarithms
and checking the scale dependence of the non-analytic portions of the
chiral loops by varying the scale.
We will assume that the divergences 
have been absorbed by a redefinition of the LECs and use
the same notation for the renormalized coupling constants. 
In the present calculation
this amounts to keeping only the chiral logarithms of the loops with the
Goldstone bosons.
(A more rigorous investigation of renormalization is provided within a modified framework
in the subsequent section.
The advantage of this approach is that the
complete renormalization of the one-loop function can be performed.)

The effective Lagrangian at one-loop order quadratic in the fields $\eta$
and $\eta'$ has the form
\begin{eqnarray}  \label{eq:quadr}
\Lagr &=& 
\sfrac{1}{2} \partial_\mu \eta \Big[ 1 + T_{88}^{(2)} \Big]\partial^\mu \eta
- \sfrac{1}{2} \eta \Big[ M_{88}^{(2)}- \big(M_{08}^{(2)}\big)^2/M_{00}^{(0)}
 + M_{88}^{(4)} \Big] \eta 
\nonumber \\&&\mathord{}
+\sfrac{1}{2} \partial_\mu \eta' \Big[ 1 + T_{00}^{(4)} - 
 \sfrac{3}{4}\big( K_{00}^{(2)}\big)^2-\big( K_{08}^{(2)}\big)^2 
+\big(M_{08}^{(2)}\big/M_{00}^{(0)}\big)^2
\Big]\partial^\mu \eta'
\nonumber \\&&\mathord{}
-  \sfrac{1}{2} \eta' \Big[ M_{00}^{(0)}+ M_{00}^{(2)} 
+ M_{00}^{(4)}
+2\big(M_{08}^{(2)}\big)^2/M_{00}^{(0)}
\nonumber \\&&\mathord{}\qquad
 - K_{00}^{(2)}\big(M_{00}^{(0)}+ M_{00}^{(2)} \big) - 2 K_{08}^{(2)} M_{08}^{(2)}
+\sfrac{1}{4}\big( K_{00}^{(2)}\big)^2 M_{00}^{(0)}
\Big] \eta' .
\end{eqnarray}
We have not shown the off-diagonal elements proportional to $\eta \eta'$ since
these do not contribute to the masses at fourth chiral order.
The term $T_{88}^{(2)}$ is the fourth order correction arising from one-loop
diagrams with a $V_1(0)$ vertex and contact terms from $\Lagr^{(4)}$ in
Eq. \eqref{eq:l4}
\begin{equation} 
T_{88}^{(2)} = \frac{1}{\decay^2} \big( 24 \cbeta{4}{0} \mmean
+ 8\cbeta{5}{0} \mcirc{\eta} - \tad{K} \big)
\end{equation}
with $\tad{\phi} = \mcirc{\phi}/(16 \pi^2) \ln (\mcirc{\phi}/\mu^2)$ and
$\mu$ the scale introduced in infrared regularization.
In order to account for all contributions to the masses 
at fourth chiral order, 
$T_{00}^{(4)}$ must include two-loop diagrams with vertices from
$\Lagr^{(2)}$, one-loop graphs from $\Lagr^{(4)}$ and contact terms from
$\Lagr^{(6)}$.
Possible two-loop diagrams are the sunset diagram and double tadpoles. 
It turns out that they do not contribute to the order we are working if
infrared regularization is employed. The only contributions to
$T_{00}^{(4)}$ arise from contact terms of $\Lagr^{(6)}$ and from
one-loop diagrams -- tadpoles in our case -- with
$\Lagr^{(4)}$ vertices.
An enumeration of all possible counterterms in $\Lagr^{(6)}$ is beyond the
scope of the present investigation. 
We will only need terms proportional to $\cder^\mu\eta_0\cder_\mu\eta_0$,
multiplied by chirally invariant combinations of two quark mass matrices.
Setting $U=1$
the only two independent combinations are $\langle\chi\rangle^2$ and 
$\langle\chi^2\rangle$, and we summarize all contributing terms to 
$T_{00}^{(4)}$ in the following Lagrangian
\begin{equation}
\Lagr^{(6)}=\ldots+\frac{1}{2\decay^2}\Big[\gamma_1(\mmean)^2
+\gamma_2\big(2(\mmean)^2+(\msplit)^2\big)\Big]
\cder_\mu\eta_0\cder^\mu\eta_0+\ldots .
\end{equation}
Including these terms the results for $T_{00}^{(4)}$, 
$M_{88}^{(4)}$ and $M_{00}^{(4)}$ read
\begin{eqnarray} 
T_{00}^{(4)} &=& \frac{4}{\decay^4} \bigl(
2\cbeta{0}{0}+ 4 \cbeta{1}{0} + \cbeta{2}{0} +
2 \cbeta{3}{0} - \sfrac{2}{3} \beta_{4,5,17,18}
-3 \cbeta{13}{0} - 6 \cbeta{14}{0} - \sfrac{3}{2} \cbeta{15}{0}\bigr)
\nonumber \\&& \qquad  
\cdot\big( 3 \mcirc{\pi} \tad{\pi} + 4 \mcirc{K} \tad{K}  + \mcirc{\eta} \tad{\eta}\big)
+\frac{1}{\decay^2}\big(\gamma_1(\mmean)^2+\gamma_2(\msplit)^2\big).
\nonumber \\
M_{88}^{(4)} &=& 
\frac{1}{\decay^2} \big( -\sfrac{1}{2} \mcirc{\pi} \tad{\pi} 
+\sfrac{1}{3} \mcirc{\pi} \tad{K} -\sfrac{8}{9}  \mcirc{K} \tad{\eta} 
+\sfrac{7}{18} \mcirc{\pi} \tad{\eta} \big) 
\nonumber \\&&\mathord{}
+ \frac{8}{\decay^2} \big( 6\cbeta{6}{0} \mcirc{\eta}\mmean+3\cbeta{7}{0}(\msplit)^2 
+ 2\cbeta{8}{0}(\mcirc{\eta})^2+\cbeta{8}{0}(\msplit)^2 \big)  
\nonumber \\
M_{00}^{(4)} &=& \frac{4}{3\decay^4} \big(-\cvtwid{2}{2}-3\coeffv{1}{2}\big) 
\big( 3 \mcirc{\pi} \tad{\pi}+ 4 \mcirc{K} \tad{K} + \mcirc{\eta} \tad{\eta} \big) 
\nonumber \\&&\mathord{}
+ \frac{24}{\decay^2} \big(2 \cbeta{6}{0}-\sqrt{6}\cbeta{26}{1}
-3\cbeta{6}{2}+ 2 \cbeta{7}{0} \big)
(\mmean)^2 
\nonumber \\&&\mathord{}
+\frac{1}{\decay^2}\big(8\cbeta{8}{0}-4\sqrt{6}\cbeta{25}{1}-6\cbeta{8}{2}-3\cbeta{12}{2}\big)
\big(2(\mmean)^2+(\msplit)^2\big).
\end{eqnarray}
The other terms in Eq. \eqref{eq:quadr} are higher order
corrections from the transformation in Eq. \eqref{eq:waverenormlag} which were not
presented for the calculation at second chiral order.

The transformations
\begin{eqnarray}\label{eq:renormcorr}
\eta  &\rightarrow&  \Big[ 1 - \sfrac{1}{2} T_{88}^{(2)}\Big] \eta \nonumber \\
\eta'  &\rightarrow&  \Big[ 1 - \sfrac{1}{2} T_{00}^{(4)} +
  \sfrac{3}{8} \big( K_{00}^{(2)}\big)^2 + \big( K_{08}^{(2)}\big)^2 
  -\sfrac{1}{2}\big(M_{08}^{(2)}\big/M_{00}^{(0)}\big)^2\Big] \eta' 
\end{eqnarray}
bring the diagonal elements of
the kinetic terms into the canonical form and change the masses
\begin{eqnarray}   \label{eq:massfourformal}
m_\eta^2  &=&  M_{88}^{(2)}+ M_{88}^{(4)}
-T_{88}^{(2)} M_{88}^{(2)} - \big(M_{08}^{(2)}\big)^2\big/M_{00}^{(2)}
\nonumber \\
m_{\eta'}^2  &=&  
M_{00}^{(0)}+M_{00}^{(2)} + M_{00}^{(4)}
- K_{00}^{(2)}\big(M_{00}^{(0)}+ M_{00}^{(2)}-M_{00}^{(0)}K_{00}^{(2)}\big) 
- T_{00}^{(4)}M_{00}^{(0)}
\nonumber \\&& \mathord{}
+\big(M_{08}^{(2)}-M_{00}^{(0)} K_{08}^{(2)}\big)^2\big/M_{00}^{(0)}.
\end{eqnarray}
Substitution of these terms gives
\begin{eqnarray}\label{eq:massfour}
m^2_{\pi}&=&\mcirc{\pi}\bigg[1
+8\big(2\cbeta{8}{0}-\cbeta{5}{0}\big)\frac{\mcirc{\pi}}{\decay^2}
+24\big(2\cbeta{6}{0}-\cbeta{4}{0}\big)\frac{\mmean}{\decay^2}
+\frac{\sfrac{1}{2}\tad{\pi}-\sfrac{1}{6}\tad{\eta}}{\decay^2}\bigg],
\nonumber\\
m^2_{K}&=&\mcirc{K}\bigg[1
+8\big(2\cbeta{8}{0}-\cbeta{5}{0}\big)\frac{\mcirc{K}}{\decay^2}
+24\big(2\cbeta{6}{0}-\cbeta{4}{0}\big)\frac{\mmean}{\decay^2}
+\frac{\sfrac{1}{3}\tad{\eta}}{\decay^2}\bigg],
\nonumber\\
m^2_{\eta}&=&\mcirc{\eta}\bigg[1
+8\big(2\cbeta{8}{0}-\cbeta{5}{0}\big)\frac{\mcirc{\eta}}{\decay^2}
+24\big(2\cbeta{6}{0}-\cbeta{4}{0}\big)\frac{\mmean}{\decay^2}\bigg]
\nonumber\\&&\mathord{}
+\frac{\big(\msplit\big)^2}{\decay^2}
 \bigg[8\cbeta{8}{0}+24\cbeta{7}{0}
-\frac{8\big(\cvtwid{2}{1}\big)^2}{\decay^2\mtop}\bigg]
\nonumber\\&&\mathord{}
+\frac{-\sfrac{1}{2}\mcirc{\pi}\tad{\pi}+\sfrac{4}{3}\mcirc{K}\tad{K}
 +\sfrac{7}{18}\mcirc{\pi}\tad{\eta}-\sfrac{8}{9}\mcirc{K}\tad{\eta}}{\decay^2},
\nonumber\\
m^2_{\eta'}&=&\mcirc{\eta'}
+\frac{8\mmean\big(\mtop-\mcirc{\eta'}\big)\beta_{4,5,17,18}}{\decay^2}
+\frac{8\big(\msplit\big)^2\bigl(\cvtwid{2}{1}-2\mtop\beta_{5,18}\bigr)^2}
{\decay^4\mtop}
\nonumber\\&&\mathord{}
+\frac{3\mcirc{\pi}\tad{\pi}+4\mcirc{K}\tad{K}+\mcirc{\eta}\tad{\eta}}{\decay^4}
\Big[-\sfrac{4}{3}\cvtwid{2}{2}-4\coeffv{1}{2}+\sfrac{8}{3}\mtop\beta_{4,5,17,18}
\nonumber\\&&\mathord{}\qquad\quad
+\mtop\big(-8\cbeta{0}{0}-16\cbeta{1}{0}-4\cbeta{2}{0}-8\cbeta{3}{0}
  +12\cbeta{13}{0}+24\cbeta{14}{0}+6\cbeta{15}{0}\big)\Big]
\nonumber\\&&\mathord{}
+\frac{\big(\mmean\big)^2}{\decay^2}
\bigl(48\cbeta{6}{0}-24\sqrt{6}\cbeta{26}{1}-72\cbeta{6}{2}+48\cbeta{7}{0}
-\mtop\gamma_1\bigr)
\nonumber\\&&\mathord{}
+\frac{2\big(\mmean\big)^2+\big(\msplit\big)^2}{\decay^2}
\bigl(8\cbeta{8}{0}-4\sqrt{6}\cbeta{25}{1}-6\cbeta{8}{2}-3\cbeta{12}{2}-\mtop\gamma_2\bigr)
\end{eqnarray}
This completes the calculation of the $\eta$ and $\eta'$ masses up to
fourth chiral order. In \cite{BW} it was assumed that the $\eta$-$\eta'$ 
mixing follows the one-mixing angle scheme and some of the terms 
in Eq. \eqref{eq:massfourformal} have been neglected. 
(This was sufficient in order to
establish infrared regularized $U(3)$ chiral perturbation theory, and the main
purpose of this paper was to show that the chiral series for the masses and
decay constants converge faster than in the dimensionally regularized theory.)
A rigorous treatment of the masses up to fourth chiral order, however,
requires the transformation
$(\eta_8,\eta_0)^T=(1+R^{(2)}+R^{(4)})(\eta,\eta')^T$ with
\begin{equation}\label{eq:lastmix}
1+R^{(2)}=\left(\begin{array}{cc}
1-\frac{1}{2}T_{88}^{(2)}&M_{08}^{(2)}\big/M_{00}^{(0)}-K_{08}^{(2)}\\[3pt]
-M_{08}^{(2)}\big/M_{00}^{(0)}
&1-\frac{1}{2} K_{00}^{(2)}\end{array}\right).
\end{equation}
where we have presented only the terms up to second chiral order for brevity.
(The fourth order terms only give contributions to $m_{\eta'}$
and scattering processes involving several $\eta'$.)
This generalizes Eq. \eqref{eq:waverenormlag},
and the entries of $R^{(2)}$ are given by
\begin{eqnarray}
R^{(2)}_{\pi}&=&\bigl(-12\mmean\cbeta{4}{0}
-4\mcirc{\pi}\cbeta{5}{0}
+\sfrac{1}{3}\tad{\pi}+\sfrac{1}{6}\tad{K}\bigr)/\decay^2,
\nonumber\\
R^{(2)}_{K}&=&\bigl(-12\mmean\cbeta{4}{0}-4\mcirc{K}\cbeta{5}{0}
+\sfrac{1}{8}\tad{\pi}+\sfrac{1}{4}\tad{K}+\sfrac{1}{8}\tad{\eta}\bigr)
/\decay^2,
\nonumber\\
R^{(2)}_{8\eta}&=&\bigl(-12\mmean\cbeta{4}{0}-4\mcirc{\eta}\cbeta{5}{0}
+\sfrac{1}{2}\tad{K}\bigr)/\decay^2,
\nonumber\\
R^{(2)}_{8\eta'}&=&2\sqrt{2}\msplit
\bigl(2\mtop\beta_{5,18}-\cvtwid{2}{1}\bigr)\big/\decay^2\mtop,
\nonumber\\
R^{(2)}_{0\eta}&=&2\sqrt{2}\msplit \cvtwid{2}{1}\big/\decay^2\mtop,
\nonumber\\
R^{(2)}_{0\eta'}&=&-4\mmean\beta_{4,5,17,18}/\decay^2,
\end{eqnarray}
where the expressions for the pions and kaons  have been included for 
completeness. The matrix $R^{(2)}$ 
constitutes one of our main results since it will be needed in
all one-loop calculations.

\subsection{Renormalization}
From the above formulas it becomes apparent that $\eta'$ loops do not contribute
at this order in infrared regularization. The tadpole which is the only one-loop
graph in the present investigation vanishes in the case of the $\eta'$.
A similar observation is made in \cite{BB} where both tadpoles and chiral unitarity
corrections have been evaluated for the hadronic decay $\eta' \rightarrow
\eta \pi \pi$. Employing infrared regularization loops with an $\eta'$
contribute at higher orders than pure Goldstone boson loops. For the processes
considered so far the infrared physics stemming from the propagation of an
$\eta'$ inside the loop is suppressed by one chiral order and therefore beyond
the working accuracy of \cite{BB} and the present investigation.
At this order it is therefore equivalent to a scheme in which the $\eta'$
is not taken into account at all in loops but rather treated first as 
a background field.
Only Goldstone boson loops occur within this approach and they
are calculated employing dimensional regularization.
After the evaluation of the loops the $\eta'$ field can be dealt with as
a propagating field.
The main advantage of such a framework is given by the complete renormalization
of the one-loop functional which cannot be undertaken in infrared regularization
since it involves the renormalization of counterterms of infinite order.
In addition to being an alternative approach for describing $\eta'$ physics at
low energies it provides a check on the renormalization of the Goldstone boson
integrals in infrared regularization.

In the appendices, we present a list of all operators of the fourth order
Lagrangian and the complete renormalization of the one-loop functional of the Goldstone boson loops.
We would like to point out that our results for the renormalization
differ substantially from those in \cite{H-S} since within this work the authors
treated the $\eta'$ on the same footing as the Goldstone bosons and included
the $\eta'$ inside loops.

\section{Results}\label{sec:results}

The decay constant $\decay$ is taken to be $88\MeV$,
the value of the pion decay constant in the chiral limit \cite{GerL}. The
quark mass matrix is chosen to fit $\mcirc[]{\pi}=138\MeV$ and
$\mcirc[]{K}=496\MeV$. We take the values of 
$\cbeta{i}{0}=\beta^{SU(3)}_{i}$ from ordinary $SU(3)$ chiral perturbation 
theory
\cite{BEG} unless stated otherwise.

\subsection{Masses}

First we investigate the mass of the $\eta'$ at second chiral order in
Eq. \eqref{eq:masstwo}
\begin{equation}
\mcirc{\eta'} =
\mtop + \frac{4}{\decay^2} \mmean
\big(\cvtwid{2}{2}-2\mtop\beta_{4,5,17,18} \big).
\end{equation}
The phenomenological values for $\cbeta{17}{0}$ and $\cbeta{18}{0}$ are
not known, but they are
OZI violating corrections to $\cbeta{5}{0}$.
Assuming that they are suppressed at the scale $\mu=m_{\rho}$, i.e. 
$|\cbeta{17}{0}(m_{\rho})|,|\cbeta{18}{0}(m_{\rho})|\ll 
|\cbeta{5}{0}(m_{\rho})|$, 
we can roughly estimate $\beta_{4,5,17,18}\approx 0.5\times 10^{-3}$.
However, this result is very sensitive to the scale at which
the OZI rule has been applied: at $\mu=m_{\eta}$ it yields
$\beta_{4,5,17,18}\approx 2.0\times 10^{-3}$. 

In order to obtain a bound for $\beta_{4,5,17,18}$ we will 
consider the dependence of $\mcirc{\eta'}$ on
$\mtop$. 
We assume that the proportionality factor 
$(1-8\beta_{4,5,17,18}\mmean/\decay^2)$ converges reasonably fast,
i.e., that the second term is at most
a $25\%$ correction to the leading order. This gives
the limit $|\beta_{4,5,17,18}|<1.5\times 10^{-3}$.

Next we assume $0<\cvtwid{2}{2}<\sfrac{1}{2}\decay^2$
in agreement with large $N_c$ considerations
and solve $\mcirc[]{\eta'}=958\MeV$ for $\mtop[]$. Under the above
assumptions this is possible only if $650\MeV<\mtop[]<1.1\GeV$.
These bounds agree with those found in \cite{GOG}.
Using the value for the topological susceptibility given within this work
which corresponds to $\coeffv{0}{2}=0.003174\GeV^4$ in our framework
we obtain
$\mtop[]=905\MeV$ ($\mtop=857\MeV$) for $\decay=88\MeV$ ($\decay=93\MeV$).

The masses for the octet from Eq. \eqref{eq:massfour} are exactly the
same as in $SU(3)$ perturbation theory
provided the LECs are related by $\cbeta{k}{0}=\beta^{SU(3)}_k$ for
$k=4,5,6,8$ and 
\begin{equation}\label{eq:beta7sat}
\beta^{SU(3)}_{7} =\cbeta{7}{0}
-\frac{\bigl(\cvtwid{2}{1}\bigr)^2}{3\decay^2\mtop}.
\end{equation}
This is in agreement with the results from \cite{H-S3,GL}.
In \cite{H-S3}
the $\eta'$ field was integrated out explicitly
to match the LECs to their $SU(3)$ values.
Note that within the approach of large $N_c$ chiral perturbation theory 
$\coeffv{3}{1}$ is of higher order and 
does not appear at the order considered there.
The phenomenological value of $\beta_7^{SU(3)}=(-0.35\pm 0.2)\times 10^{-3}$ 
may be saturated completely by the additional term in
Eq. \ref{eq:beta7sat}.

\subsection{Decay constants}

Phenomenologically the $\eta$-$\eta'$ mixing can be extracted from
the pseudoscalar decays. 
The decay constants $F_{kl}$ are defined by the processes 
$\langle 0|A^{l}_\mu|\phi_k\rangle=i p_\mu F_{kl}$.
At lowest order the decay constants
are $F_{kl}=\decay\delta_{kl}$ for the octet and
\begin{equation} \label{eq:decay0}
F_{\eta'0}=\decay_{0}=\frac{\sqrt{6\lambda}\big(\decay^2+6\coeffv{5}{0}\big)}{\decay^2}
\end{equation}
for the singlet.
At next-to-leading order there are also off-diagonal 
decay constants where mixing effects appear.
The mixing will be parametrized by 
\begin{eqnarray}
(F_{\eta8},F_{\eta'8})&=&F_8(\cos \vartheta_8,\sin\vartheta_8),\nonumber
\\
(F_{\eta0},F_{\eta'0})&=&F_0(-\sin \vartheta_0,\cos\vartheta_0),
\end{eqnarray}
while no other mixing occurs among the decay constants
in the isospin limit $m_u=m_d$. 

In this section we need a few more operators from 
the fourth order Lagrangian \eqref{eq:l4}, namely $O_k$ with $k=46,47,52,53$.
We find the decay constants at next-to-leading order 
in a similar way as the masses in Sec.~\ref{sec:loopmass}:
\begin{eqnarray}\label{eq:axcouptwo}
F_\pi&=&\decay\Big[1+12 \cbeta{4}{0} \frac{\mmean}{\decay^2}
+4 \cbeta{5}{0} \frac{\mcirc{\pi}}{\decay^2}
-\frac{\tad{\pi}+\sfrac{1}{2}\tad{K}}{\decay^2}\Big],
\nonumber\\
F_K&=&\decay\Big[1+12 \cbeta{4}{0} \frac{\mmean}{\decay^2}
+4 \cbeta{5}{0} \frac{\mcirc{K}}{\decay^2}
-\frac{\sfrac{3}{8}\tad{\pi}+\sfrac{3}{4}\tad{K}+\sfrac{3}{8}\tad{\eta}}{\decay^2}\Big],
\nonumber\\
F_{\eta8}&=&\decay\Big[1+12 \cbeta{4}{0} \frac{\mmean}{\decay^2}
+4 \cbeta{5}{0} \frac{\mcirc{\eta}}{\decay^2}
-\frac{\sfrac{3}{2}\tad{K}}{\decay^2}\Big],
\nonumber\\
F_{\eta'8}&=&-\frac{2\sqrt{2}\msplit\cvtwid{2}{1}}{\mtop \decay},
\nonumber\\
F_{\eta0}&=&\frac{2\sqrt{2}\msplit}{\decay^2}
\bigg[\frac{\decay_0\cvtwid{2}{1}}{\mtop}
-\sqrt{6\lambda}\bigl(2\beta_{5,18}
+3\cbeta{46}{0}+3\cbeta{53}{0}\bigr)\bigg],
\nonumber\\
F_{\eta'0}&=&\decay_0
+\frac{4\mmean(2\sqrt{6\lambda}-\decay_0)\beta_{4,5,17,18}}{\decay^2}
\nonumber\\&&\qquad\mathrel{}
+\frac{\sqrt{6\lambda}\mmean}{\decay^2}
\big(12\cbeta{46}{0}+36\cbeta{47}{0}-12\cbeta{53}{0}-6\sqrt{6}\cbeta{52}{1}\bigr).
\end{eqnarray}
In second order the two decay amplitudes $F_8,F_0$ are given by 
$F_8=F_{\eta 8}$ and $F_0=F_{\eta' 0}$, while the
angles $\vartheta_8,\vartheta_0$ are
\begin{eqnarray}\label{eq:axangletwo}
\vartheta_8
&=&-\frac{2\sqrt{2}\msplit}{\mtop \decay^2}\cvtwid{2}{1},
\nonumber\\
\vartheta_0
&=&\vartheta_8
+\frac{2\sqrt{2}\msplit}{\decay^2+6\coeffv{5}{0}}
\bigl(2\beta_{5,18}
+3\cbeta{46}{0}+3\cbeta{53}{0}\bigr).
\end{eqnarray}
This is the leading order contribution to the mixing angles and 
both angles differ. 
Phenomenological values for the angles have been given, e.g. in \cite{FK2}:
$\vartheta_8=-21.2^\circ$, $\vartheta_0=-9.2^\circ$.

\subsection{Fit}

We will use the above equations to fit some of the
parameters. To be more precise, we use the mass formula of 
the $\eta'$ at second chiral order, the mixing angle $\vartheta_8$
and assume the complete saturation of $\beta_7^{SU(3)}$ 
due to $\eta$-$\eta'$ mixing (i.e. $\beta_7^{(0)} \approx 0$) 
in order to obtain values for
the parameters $\mtop$, $\cvtwid{2}{1}$ and $\beta_{4,5,17,18}$.
The $1/N_c$ estimate for $\cvtwid{2}{2}$ reads 
$\cvtwid{2}{2} \approx 2\cvtwid{2}{1}-\sfrac{1}{4}\decay^2$.
Taking the values 
$\mcirc[]{\eta'}=958\MeV$ and 
$\vartheta_8=-21.2^\circ$
the resulting parameters are
\begin{equation} \label{eq:centralfit}
\mtop[]=847\MeV,\quad \cvtwid{2}{1}=1.25\times \sfrac{1}{4}\decay^2,
\quad \beta_{4,5,17,18}=0.47\times 10^{-3}.
\end{equation}
These values are all in the expected ranges, however, they
depend heavily on our assumptions:
A change in $\mcirc[]{\eta'}$ by $10\MeV$, e.g., requires $\beta_{4,5,17,18}$ to
change by $-0.15\times 10^{-3}$, whereas 
a change in $\vartheta_8$ by $1^\circ$ or 
in $\beta_7$ by $0.05\times 10^{-3}$ results in the changes 
\begin{eqnarray}
&\mtop[]: +\phantom{0}6\MeV,\quad \cvtwid{2}{1}:+0.10\times\sfrac{1}{4}\decay^2,\quad
\beta_{4,5,17,18}:+0.33\times 10^{-3},&
\nonumber\\
&\mtop[]: +59\MeV,\quad \cvtwid{2}{1}:+0.18\times \sfrac{1}{4}\decay^2,\quad
\beta_{4,5,17,18}:+1.08\times 10^{-3}.&
\end{eqnarray}
The value for $\mtop[]$ is in agreement with 
the result given in \cite{GOG}.

At fourth chiral order the evaluation of the $\eta'$ mass is rendered
more difficult due to the proliferation 
of new counterterms. We will therefore make
the following rough estimate by neglecting the unknown OZI violating
couplings and keeping only the known parameters.
The terms of fourth chiral order for $m_{\eta'}$ in Eq.~\eqref{eq:massfour}
are then---in order of appearance---corrections of about
$-1\%$, $2.5\%$, $40\%$, $2\%$, $7\%$ relatively to $\mcirc[]{\eta'}$.
The loop term delivers by far the greatest contribution but is highly
scale dependent. This can be seen immediately, e.g., by noting that
the prefactor 
$3\mcirc{\pi}\tad{\pi}+4\mcirc{K}\tad{K}+\mcirc{\eta}\tad{\eta}$
vanishes at a scale of about $\mu = 520\MeV$. The counterterms included
in $\gamma_1$ and $\gamma_2$ which cancel this scale dependence will therefore
also vary strongly with $\mu$ and might lead to sizeable contributions
depending on the choice for $\mu$. In order to confine their approximate
size one must consider further processes involving these couplings.

\section{Conclusions}
\label{sec:conclusions}

In this investigation we have presented  $\eta$-$\eta'$ mixing up to
one-loop
order in the context of the masses and decay constants of the
$\eta$-$\eta'$ 
system. We worked in the framework of infrared regularized $U(3)$ chiral
perturbation theory which permits a strict chiral counting scheme without
employing large $N_c$ counting rules. 
We treat the $\eta'$ as a massive state, whereas it is considered
to be a small quantity in large $N_c$ chiral perturbation theory.
It turns out that even at leading
order the $\eta$ and $\eta'$ fields do not follow the usually assumed
one-mixing angle scheme. Already at tree level the mixing of these states
cannot be parametrized by just one single angle
which is in contradistinction to large $N_c$ chiral perturbation theory
where one mixing angle is sufficient at lowest order.
In this framework the physical fields $\eta$ and $\eta'$ are related to the pure 
octet and singlet states, $\eta_8$ and $\eta_0$, via a matrix which includes at 
leading order the parameter combinations $\cvtwid{2}{1}, \beta_{5,18}$
and $\beta_{4,5,17,18}$ as well as $\mtop[]$, the
$\eta'$ mass  in the chiral limit of vanishing quark masses, see Eq. 
\eqref{eq:waverenormlag}.
As an immediate consequence, matrix elements involving $\eta$ and $\eta'$
fields will include these parameter combinations  and can be used to extract their
numerical values by comparison with experimental data. The pseudoscalar decays, 
e.g., are suited to obtain reasonable estimates for the couplings and a fit
to the two angles $\vartheta_8$ and $\vartheta_0$ can be easily accomodated 
as shown in the preceding section. However, using the results from Eq. 
\eqref{eq:axangletwo}, we can turn the argument around and obtain a rough estimate
for $\vartheta_8$ and $\vartheta_0$. To this end, we assume that the values of the
1/$N_c$ suppressed couplings are negligible, i.e. in particular  
$|\coeffv{5}{0}| \ll f^2/4$, and $|\beta_{18}^{(0)}|, |\beta_{46}^{(0)}|, 
|\beta_{53}^{(0)}| \ll |\beta_{5}^{(0)}|$, hence generalizing our approximation
for the OZI violating contributions made in the last section.
The parameter $\beta_{5}^{(0)}$ itself is phenomenologically determined
by the ratio \cite{GL,BEG}
\begin{equation}
\frac{F_K}{F_\pi} =  1 + 4 ( m_K^2 -m_\pi^2) \frac{\beta_{5}^{(0)}}{\decay^2}
+ \frac{5}{8 \decay^2} \tad{\pi} - \frac{1}{4 \decay^2} \tad{K}
- \frac{3}{8 \decay^2} \tad{\eta} \approx  1.22 .
\end{equation}
Using $\decay =88\MeV$ yields the value $\beta_{5}^{(0)} = 1.3\times 10^{-3}$ which is
consistent with \cite{BEG}.
With these rough assumptions we obtain
\begin{equation} \label{eq:diffangle}
\vartheta_0 - \vartheta_8
=
\frac{16 \sqrt{2}(m_K^2 - m_\pi^2)}{3 \decay^2} \beta_{5}^{(0)} = 16.4^\circ.
\end{equation}
which slightly overestimates, e.g., the phenomenological 
extraction of \cite{FK2}.
This indicates that other contributions such as the neglected LECs from Eq. 
\eqref{eq:axangletwo} or higher orders may modify our estimate
for $\vartheta_0 - \vartheta_8$; nevertheless,
the two angles differ considerably.
The result is similar to the one obtained in \cite{KL1}.
Note, however, that within the present scheme, this is the leading contribution, while 
in the combined chiral and 1/$N_c$ expansion the difference
of the two angles starts at subleading order and the form as given in Eq. 
\eqref{eq:diffangle} corresponds even to next-to-next-to-leading order.

The phenomenological determination of $\eta$-$\eta'$ mixing from photonic decays
of the $\eta$ and $\eta'$ should yield a more reliable value. The lowest order 
contribution to these decays originates from the anomalous Wess-Zumino-Witten
term which is of fourth chiral order.
In order to pin down the values of the two angles accurately, one must    
calculate $SU(3)$ breaking corrections to the Wess-Zumino-Witten term which
are of sixth chiral order and beyond the scope of the present investigation.

Under reasonable assumptions for the parameters of the $\eta'$ mass at
second chiral order we were able to obtain a range for $\mtop[]$: 
$650\MeV<\mtop[]<1.1\GeV$, i.e. in our approach it is in principle possible
that the $\eta'$ mass contribution due to the axial $U(1)$ anomaly can be
larger than the physical mass of $958\MeV$ and is lowered by leading order
symmetry breaking terms. 
Comparing the mixing angle $\vartheta_8$ with phenomenological analyses, 
and assuming that $\beta_7^{SU(3)}$ is completely saturated by
the $\eta'$ resonance, we were able to disentangle two of the parameters:
$\cvtwid{2}{1}$, which is predominantly
responsible for $\eta$-$\eta'$ mixing and 
$\mtop[] \approx 850\MeV$.
This value for $\mtop[]$ is in agreement with other analyses 
(see e.g. \cite{GOG}) and
it shows that the saturation of $\beta_7^{SU(3)}$ was a consistent assumption.

The mass $\mtop$ is given by $\mtop = 2 \coeffv{0}{2}/ \decay^2$, see
Eq. \eqref{eq:abbrev}, a well-known result \cite{WV}.
In the large $N_c$ limit $\coeffv{0}{2}$ coincides with $3 \tau_{\mathrm{\scriptscriptstyle{GD}}}$, where
$\tau_{\mathrm{\scriptscriptstyle{GD}}}$ is the topological susceptibility of Gluodynamics. It represents
the mean square winding number per unit volume of euclidean space
\begin{equation}
\tau_{\mathrm{\scriptscriptstyle{GD}}} \equiv  \int d^4 x \langle 0| T \omega (x) \omega (0) | 0
\rangle_{\mathrm{\scriptscriptstyle{GD}}}
\end{equation}
with
\begin{equation}
\omega = \frac{g^2}{16 \pi^2} \: \mbox{tr} \, G_{\mu \nu} \tilde{G}^{\mu \nu} .
\end{equation}
The uncertainty in $\mtop[]$ translates immediately into a range for 
$\tau_{\mathrm{\scriptscriptstyle{GD}}}$
\begin{equation}
0.55 \times 10^{-3} \GeV^4 < \tau_{\mathrm{\scriptscriptstyle{GD}}} < 1.56 \times 10^{-3} \GeV^4 .
\end{equation}

However, some of the results are rather sensitive to the assumptions
made for the parameters. A further study of the $\eta$-$\eta'$ system, such as their
hadronic decay modes as well as the anomalous decays, should yield more
reliable values for some parameters and the mixing angles as they appear
in the parametrization of the pseudoscalar decay constants for the $\eta$
and $\eta'$ \cite{BB}.

\section*{Acknowledgements}

We would like to thank Stefan Wetzel for reading the manuscript.

\appendix

\section{Fourth order operators}\label{sec:fourth}

We use the standard definitions of chiral perturbation theory
\begin{eqnarray}
F^{L}_{\mu\nu}&=&\partial_\mu \tilde l_{\nu}-\partial_\nu \tilde l_{\mu}-i [\tilde l_\mu,\tilde l_\nu],
\nonumber\\
F^{R}_{\mu\nu}&=&\partial_\mu \tilde r_{\nu}-\partial_\nu \tilde r_{\mu}-i [\tilde r_\mu,\tilde r_\nu],
\nonumber\\
\chi&=&2B(s+ip),
\end{eqnarray}
where $U$ is a unitary matrix containing the meson fields.
The fields $s$, $p$, $v_\mu=\sfrac{1}{2}(\tilde r_\mu+\tilde l_\mu)$
and $\tilde a_\mu=\sfrac{1}{2}(\tilde r_\mu-\tilde l_\mu)$
are the external sources that couple to the QCD Lagrangian. 
The singlet axial-vector source $\langle \tilde a_\mu\rangle$ 
has been rescaled to account for the
dependence on the running QCD scale, Eq. \eqref{eq:scalea0}.

We make use of the following abbreviations for the definition of the fourth order Lagrangian
\begin{eqnarray} \label{eq:abr}
C_{\mu}&=&U^\dagger \cder_\mu U,
\nonumber\\
T_\mu&=&i \cder_\mu \theta,
\nonumber\\
M&=&U^\dagger \chi+\chi^\dagger U,
\nonumber\\
N&=&U^\dagger \chi-\chi^\dagger U,
\nonumber\\
F^\pm_{\mu\nu}&=&F^L_{\mu\nu}\pm U^\dagger F^R_{\mu\nu} U,
\end{eqnarray}
The fourth order operators $O_k$ are given in Tab. \ref{tab:fourth} and
the fourth order Lagrangian 
\begin{equation}
\Lagr^{(4)}=\tsum\nolimits_k \beta_k O_k.
\end{equation}
is a sum of these operators coupled with functions $\beta_k$ of the 
invariant $\eta_0+\sqrt{\lambda}\,\theta$, which can be 
expanded as in Eq. \eqref{eq:vexpand}.

\begin{table}
\[
\begin{array}{ll}
O_{\phantom{0}0}=\langle C^\mu C^\nu C_\mu C_\nu\rangle,&
O_{\phantom{0}1}=\langle C^\mu C_\mu\rangle\langle C^\nu C_\nu\rangle,\\
O_{\phantom{0}2}=\langle C^\mu C^\nu\rangle\langle C_\mu C_\nu\rangle,&
O_{\phantom{0}3}=\langle C^\mu C_\mu C^\nu C_\nu\rangle,\\
O_{13}=-\langle C^\mu\rangle\langle C_\mu C^\nu C_\nu\rangle,&
O_{14}=-\langle C^\mu\rangle\langle C_\mu\rangle \langle C^\nu C_\nu\rangle,\\
O_{15}=-\langle C^\mu\rangle\langle C^\nu\rangle \langle C_\mu C_\nu\rangle,\qquad&
O_{16}=\langle C^\mu\rangle\langle C_\mu\rangle\langle C^\nu\rangle\langle C_\nu\rangle,\\
O_{\phantom{0}4}=-\langle C^\mu C_\mu\rangle\langle M\rangle,&
O_{\phantom{0}5}=-\langle C^\mu C_\mu M\rangle,\\
O_{17}=\langle C^\mu \rangle\langle C_\mu\rangle\langle M\rangle,&
O_{18}=-\langle C^\mu \rangle\langle C_\mu M\rangle,\\
O_{21}=\langle C^\mu C_\mu i N\rangle,&
O_{22}=\langle C^\mu C_\mu\rangle\langle i N\rangle,\\
O_{23}=\langle C^\mu \rangle\langle C_\mu i N\rangle,&
O_{24}=\langle C^\mu \rangle\langle C_\mu\rangle\langle i N\rangle,\\
O_{\phantom{0}6}=\langle M\rangle\langle M\rangle,&
O_{\phantom{0}7}=\langle N\rangle\langle N\rangle,\\
O_{\phantom{0}8}=\sfrac{1}{2}\langle MM+NN\rangle,&
O_{12}=\sfrac{1}{4}\langle MM-NN\rangle,\\
O_{25}=\langle iMN\rangle,&
O_{26}=\langle M\rangle\langle iN\rangle,\\
O_{\phantom{0}9}=i\langle C_\mu C_\nu F_+^{\mu\nu}\rangle,&
O_{27}=\langle C_\mu \rangle\langle C_\nu F_-^{\mu\nu}\rangle,\\
O_{29}=i\varepsilon_{\mu\nu\rho\sigma}\langle C^\mu C^\nu F_+^{\rho\sigma}\rangle,&
O_{30}=\varepsilon_{\mu\nu\rho\sigma}\langle C^\mu\rangle \langle C^\nu F_-^{\rho\sigma}\rangle,\\
O_{10}=\sfrac{1}{4}\langle F_+^{\mu\nu}F^+_{\mu\nu}-F_-^{\mu\nu}F^-_{\mu\nu}\rangle,&
O_{11}=\sfrac{1}{2}\langle F_+^{\mu\nu}F^+_{\mu\nu}+F_-^{\mu\nu}F^-_{\mu\nu}\rangle,\\
O_{20}=\sfrac{1}{4}\langle F_+^{\mu\nu}\rangle\langle F^+_{\mu\nu}\rangle-\sfrac{1}{4}\langle F_-^{\mu\nu}\rangle\langle F^-_{\mu\nu}\rangle,\,\,\,&
O_{19}=\sfrac{1}{2}\langle F_+^{\mu\nu}\rangle\langle F^+_{\mu\nu}\rangle+\sfrac{1}{2}\langle F_-^{\mu\nu}\rangle\langle F^-_{\mu\nu}\rangle,\\
O_{28}=\sfrac{1}{4}\varepsilon_{\mu\nu\rho\sigma}\langle F_+^{\mu\nu}F_+^{\rho\sigma}-F_-^{\mu\nu}F_-^{\rho\sigma}\rangle,\\
O_{31}=T^\mu \langle C_\mu C^\nu C_\nu\rangle,&
O_{32}=T^\mu \langle C_\mu\rangle \langle C^\nu C_\nu\rangle,\\
O_{33}=T^\mu \langle C_\mu C^\nu\rangle\langle C_\nu\rangle,&
O_{34}=T^\mu \langle C_\mu\rangle \langle C^\nu \rangle\langle C_\nu\rangle,\\
O_{35}=T^\mu T_\mu \langle C^\nu C_\nu\rangle,&
O_{37}=T^\mu T_\mu \langle C^\nu \rangle\langle C_\nu\rangle,\\
O_{36}=T^\mu T^\nu \langle C_\mu C_\nu\rangle,&
O_{38}=T^\mu T^\nu \langle C_\mu \rangle\langle C_\nu\rangle,\\
O_{39}=T^\mu T_\mu T^\nu \langle C_\nu\rangle,&
O_{40}=T^\mu T_\mu T^\nu T_\nu,\\
O_{41}=i \cder^\mu T_\mu \langle C^\nu C_\nu \rangle,&
O_{42}=i \cder^\mu T_\mu \langle C^\nu \rangle\langle C_\nu \rangle,\\
O_{43}=i \cder^\mu T_\mu T^\nu\langle C_\nu \rangle,&
O_{44}=i \cder^\mu T_\mu T^\nu T_\nu,\\
O_{45}=\cder^\mu T_\mu  \cder^\nu T_\nu,\\
O_{46}=T^\mu\langle C_\mu M\rangle,&
O_{47}=T^\mu \langle C_\mu \rangle \langle M\rangle,\\
O_{48}=T^\mu \langle C_\mu iN\rangle,&
O_{49}=T^\mu \langle C_\mu \rangle\langle iN\rangle,\\
O_{50}=T^\mu T_\mu \langle M\rangle,&
O_{51}=T^\mu T_\mu \langle iN\rangle,\\
O_{52}=i \cder^\mu T_\mu \langle M\rangle,&
O_{53}=\cder^\mu T_\mu \langle N\rangle,\\
O_{54}=T_\mu \langle C_\nu F_-^{\mu\nu}\rangle,&
O_{55}=T_\mu \langle C_\nu \rangle \langle F_-^{\mu\nu}\rangle,\\
O_{56}=-\varepsilon_{\mu\nu\rho\sigma} T^\mu \langle C^\nu F_-^{\rho\sigma}\rangle,&
O_{57}=-\varepsilon_{\mu\nu\rho\sigma} T^\mu \langle C^\nu \rangle\langle F_-^{\rho\sigma}\rangle.
\end{array}
\]
\caption{Fourth order operators in $U(3)$ ChPT \cite{H-S2}}
\label{tab:fourth}
\end{table}

In standard $SU(3)$ ($U(3)$) chiral perturbation theory the fourth order Lagrangian consists
of all possible fourth order operators with coupling 
constants (functions in $\eta_0 + \sqrt{\lambda}\,\theta$)
not fixed by chiral symmetry. In total there are 13 (58) independent fourth order 
operators, one of which can be eliminated by the Cayley-Hamilton identity
for $n_l=3$ light flavors. 
The equation of motion for the meson fields has been used extensively
in order to eliminate operators involving the divergence
$\cder^\mu C_\mu$.

In the renormalization scheme presented in App. \ref{sec:renorm}, we will treat the $\eta_0$ 
as a background field that is not restricted by an equation of motion.
To this end, we need to 
include additional operators involving $\langle \cder^\mu C_\mu\rangle$. 
To second order the only new counterterm is proportional to $\langle \cder^\mu C_\mu\rangle$
which is a total divergence and equivalent to operators
already present in the second order Lagrangian. It can therefore be omitted.
At  fourth order, however, new operators 
must be included which have not been considered in previous approaches.
The eight additional counterterms read
\begin{equation}
\begin{array}{ll}
O_{58}=\I\langle \cder^\mu C_\mu\rangle\langle C^\nu C_\nu\rangle,\,\,\,\,&
O_{59}=\I\langle \cder^\mu C_\mu\rangle\langle C^\nu\rangle\langle C_\nu\rangle,\\
O_{60}=\I\langle \cder^\mu C_\mu\rangle\langle C^\nu\rangle T_\nu,&
O_{61}=\I\langle \cder^\mu C_\mu\rangle T^\nu  T_\nu ,\\
O_{62}=\langle \cder^\mu C_\mu\rangle \langle \cder^\nu C_\nu\rangle,&
O_{63}=\langle \cder^\mu C_\mu\rangle \cder^\nu T_\nu,\\
O_{64}=\I\langle \cder^\mu C_\mu\rangle \langle M\rangle,&
O_{65}=\langle \cder^\mu C_\mu\rangle \langle N\rangle.
\end{array}
\end{equation}
These operators are needed 
as long as the phase of $U$ which describes the singlet field is treated
as a background field. 
When subsequently the phase of $U$ is dealt with as a propagating
field, its equation of motion \cite{H-S2} 
may be used to eliminate the new operators. 
The amplitudes are then renormalizable only
on-shell, but
if one prefers to keep the new operators instead, they 
are renormalizable even if the $\eta'$ field is 
off-shell.
We have confirmed this property for a number of amplitudes.

\section{Renormalization}\label{sec:renorm}

In this section, we work out the 
renormalization of the one-loop functional of the Goldstone boson loops
proceeding along the lines of \cite{H-S2}
and using their notation. 
We will sketch the method briefly and highlight the
differences since the details can be found in 
\cite{H-S2,GL}.
The alternative treatment of the singlet field 
within our approach 
yields substantially different results. 
In the scheme of \cite{H-S2} the singlet is a quantum field,
whereas we treat it is as an external field which does not propagate
so that we can restrict ourselves to 
$SU(3)$ instead of $U(3)$ matrices and relations.
The results of this appendix can be used as a check for the infrared
regularized loop contributions in the present investigation since the
$\eta'$ does not appear inside loops at the order we are working.
The employed $SU(3)$ relations are 
\begin{eqnarray}
\lambda^a A\lambda^a&=&2\langle A\rangle-(2/n_l)A,
\nonumber\\
\langle\lambda^a A\rangle\langle\lambda^a B\rangle&=&
2\langle AB\rangle-(2/n_l)\langle A\rangle\langle B\rangle .
\nonumber
\end{eqnarray}
In the scheme of \cite{GL}, on the other hand, the singlet field
is not included explicitly but the methodology to extract the divergences
is equivalent. Omitting in the present investigation the
external singlet field contributions reproduces 
the results of \cite{GL}.

We start by introducing a background field $\bar{U} \in U(3)$ which 
obeys the equation of motion for the octet whereas its phase is arbitrary.
The matrix $U$ is decomposed as $U=\bar{U}\exp(\I\Delta)$ with
quantum fluctuations $\Delta \in SU(3)$.
The second order chiral Langrangian expanded up to two 
powers of $\Delta$ reads
\begin{eqnarray}  \label{eq:quantum}
\Lagr(U)&=&\Lagr(\bar{U})
+V_1(X) \langle D^\mu\Delta D_\mu\Delta\rangle
+V_1(X) \langle C^\mu [\Delta,D_\mu\Delta]\rangle
\nonumber\\&&\mathord{}
-\sfrac{1}{2}V_2(X)\langle \Delta^2 M\rangle
-\sfrac{1}{2}\I V_3(X)\langle \Delta^2 N\rangle,
\end{eqnarray}
with $D_\mu\Delta=\partial_\mu\Delta-\I[l_\mu,\Delta]$ and
the invariant quantity 
$X=\langle\log \bar{U}\rangle+i(\sqrt{6\lambda}/f)\theta=\I \sqrt{6}(\eta_0+\sqrt{\lambda}\,\theta)/f$.
The terms linear in $\Delta$ vanish upon using the equation of motion
and we drop the piece $\Lagr(\bar{U})$ which does not depend on the quantum
fluctuations $\Delta$.
Eq. \eqref{eq:quantum} corresponds to Eq. (21) in \cite{H-S2} when all terms
proportional to $\langle\Delta\rangle$ are neglected since
they vanish for $\Delta \in SU(3)$.
We then set $\Delta=\varphi^a\lambda^a/2\sqrt{V_1}$ to obtain
canonically normalized kinetic terms for the octet $\varphi$.
After partial integration and completion of a square the Lagrangian becomes
\begin{equation}
\Lagr=\sfrac{1}{2} d^\mu \varphi^a d_\mu \varphi^a
-\sfrac{1}{2}\varphi^a\sigma^{ab}\varphi^b.
\end{equation}
The connection $\omega$ of 
$d_\mu \varphi^a=\partial_\mu \varphi^a+\omega_\mu^{ab}\varphi^b$,
the curvature $R$ thereof 
and the mass term $\sigma$ read
\begin{eqnarray}
\omega_\mu^{ab}&=&
\sfrac{1}{2}\I\big\langle\omega_\mu[\lambda^a,\lambda^b]\big\rangle,
\qquad
R_{\mu\nu}^{ab}=
\sfrac{1}{2}\I\big\langle R_{\mu\nu}[\lambda^a,\lambda^b]\big\rangle,
\nonumber\\
\omega_\mu&=&l_\mu+\sfrac{1}{2}\I C_\mu,
\qquad\qquad
R_{\mu\nu}=\sfrac{1}{2}F_{\mu\nu}^L
+\sfrac{1}{2} U^\dagger F_{\mu\nu}^R U
-\sfrac{1}{4}\I [C_\mu,C_\nu],
\nonumber\\
\sigma^{ab}&=&
\sfrac{1}{8}\big\langle[C^\mu,\lambda^a][C_\mu,\lambda^b]\big\rangle
+\sfrac{1}{8}\big\langle (\omega_2 M+\I\omega_3 N)\{\lambda^a,\lambda^b\}\big\rangle
+\delta^{ab}S,
\nonumber\\
S&=&
-\sfrac{1}{2}\omega''_1\partial^\mu X\partial_\mu X
+\sfrac{1}{4}\omega'_1\omega'_1\partial^\mu X\partial_\mu X
-\sfrac{1}{2}\omega'_1\partial^\mu\partial_\mu X,
\end{eqnarray}
where we have supressed the bars in $\bar{U}$ and in the related quantities
of Eq. \eqref{eq:abr}.
The functions $\omega_k$, $\omega'_1$ and $\omega''_1$ are defined 
as the quotients $V_k/V_1$, $V'_1/V_1$ and $V''_1/V_1$, respectively. 
Note that the derivate $V'_i$ is defined as in \cite{H-S2} as
\begin{equation}
V'_1=\frac{\partial V_1}{\partial X}=\frac{1}{\I\sqrt{6}}\,
\frac{\partial V_1}{\partial (\eta_0/f)},
\end{equation}
in comparison to Eq. \eqref{eq:vexpand}. 
The differences to Eqs. (27-30) in \cite{H-S2} stem from 
the modified algebra. 

Taking the fields $\varphi$ as quantum fields, whereas 
$\bar{U},X$ are external fields, we calculate
the one loop effective action. The divergent piece in dimensional
regularization at $d=4$ is 
\begin{equation}
\frac{\Gamma}{(4\pi)^2(4-d)},
\qquad
\Gamma=\sfrac{1}{12} R^{\mu\nu}_{ab} R^{ba}_{\mu\nu}
+\sfrac{1}{2}\sigma_{ab}\sigma_{ba}.
\end{equation}
After some algebra we obtain $\Gamma$ expressed in terms
of the $58$ known and $8$ new fourth order operators 
(cf. App. \ref{sec:fourth})
\begin{eqnarray}  \label{eq:gamma}
\Gamma&=&
\frac{1}{48}
\big(
n_l O_0
+3O_1
+6O_2
+2n_l O_3
+12O_{13}
\big)
\nonumber\\&&\mathord{}
+\frac{1}{24}\big(
2n_l O_9-2n_l O_{10}-n_l O_{11}+O_{19}+2O_{20}
\big)
\nonumber\\&&\mathord{}
+\frac{1}{8}
\big(
-\omega_2 O_4
+n_l\omega_2 O_5
-2\omega_2 O_{18}
-n_l\omega_3 O_{21}
-\omega_3 O_{22}
+2\omega_3 O_{23}
\big)
\nonumber\\&&\mathord{}
+\frac{n_l^2+2}{16n_l^2}
\big(
\omega_2\omega_2 O_6
+\omega_3\omega_3 O_7
+2\omega_2\omega_3 O_{26}
\big)
\nonumber\\&&\mathord{}
+\frac{n_l^2-4}{16n_l}
\big(
(\omega_2\omega_2-\omega_3\omega_3)O_8
+(\omega_2\omega_2+\omega_3\omega_3)O_{12}
+2\omega_2\omega_3 O_{25}
\big)
\nonumber\\&&\mathord{}
+\frac{\omega'_1\omega'_1-2\omega''_1}{8}
\big(
n_l O_{14}+O_{16}^\ast-2n_l O_{32}+2O_{34}^\ast-n_l O_{35}+O_{37}^\ast
\big)
\nonumber\\&&\mathord{}
+\frac{(n_l^2-1)(\omega'_1\omega'_1-2\omega''_1)}{8n_l}
\big(
\omega_2 (O_{17}+2O_{47}+O_{50})
+\omega_3 (O_{24}+2O_{49}+O_{51})
\big)
\nonumber\\&&\mathord{}
+\frac{\I\omega'_1}{4}
\big(
-n_l O_{41}+O_{42}^\ast-n_l O_{58}+O_{59}^\ast
\big)
\nonumber\\&&\mathord{}
+\frac{(n_l^2-1)\I\omega'_1}{4n_l}
\big(
\omega_2 O_{52}
-\omega_3 O_{53}
+\omega_2 O_{64}
-\omega_3 O_{65}
\big)
\nonumber\\&&\mathord{}
+\frac{(n_l^2-1)(\omega'_1\omega'_1-2\omega''_1)^2}{32}
\big(
O_{16}^\ast
+4 O_{34}^\ast
+2 O_{37}^\ast
+4 O_{38}
+4 O_{39}
+O_{40}
\big)
\nonumber\\&&\mathord{}
+\frac{(n_l^2-1)\I\omega'_1(\omega'_1\omega'_1-2\omega''_1)}{8}
\big(
-O_{42}^\ast
-2 O_{43}
-O_{44}
-O_{59}^\ast
-2 O_{60}
-O_{61}
\big)
\nonumber\\&&\mathord{}
+\frac{(n_l^2-1)\omega'_1\omega'_1}{8}
\big(
O_{45}
+O_{62}
+2 O_{63}
\big)
\end{eqnarray}
The divergence of the one-loop effective action
needs to be cancelled by counterterms in the 
coupling functions of the fourth order operators.
The corresponding renormalization functions 
can be read off from the coefficients of the operators in $\Gamma$.
For convenience the operators which appear twice are
marked by $^\ast$.

The structure of $\Gamma$ equals that of standard $SU(3)$ chiral perturbation thory
if $\omega_2=1$, $\omega_3=\omega'_1=\omega''_1=0$ and 
all non-standard operators ignored. For $n_l=3$ the
Cayley-Hamilton matrix identity can be used to 
shuffle the coefficient of $O_0$ to those of 
$O_k$ with $k=1,2,3$ (and $13,14,15,16$).
The result for  $\Gamma$ has been confirmed by calculating 
four point amplitudes such as  $\eta' \eta' \rightarrow \eta'  \eta' $ 
scattering. After performing the renormalization prescription as given by
Eq. \eqref{eq:gamma} the amplitudes were rendered finite and independent of the
scale $\mu$ introduced in dimensional regularization.


\end{document}